\documentclass[fleqn,usenatbib]{mnras}
\usepackage{newtxtext,newtxmath}

\usepackage[T1]{fontenc}

\DeclareRobustCommand{\VAN}[3]{#2}
\let\VANthebibliography\thebibliography
\def\thebibliography{\DeclareRobustCommand{\VAN}[3]{##3}\VANthebibliography}

\usepackage{graphicx}	
\usepackage{amsmath}	
\usepackage{multirow}
\usepackage{lineno}
\usepackage{caption}
\usepackage{threeparttable}
\usepackage{gensymb}

\title[HSP Blazar IXPE Population Study]{Testing the Correlations between X-ray Spectral and Polarization Properties for High Synchrotron Peaked Blazars}

\author[M. Lynne Saade et al.]{
M. Lynne Saade,$^{1,2,3}$\thanks{E-mail: mlsaade@usra.edu)}
Steven R. Ehlert,$^{2}$
Ioannis Liodakis,$^{3}$
Philip Kaaret,$^{2}$
Fabrizio Tavecchio,$^{4}$
\newauthor
Sara Capecchiacci,$^{3,5}$
and Riccardo Middei,$^{6}$
\\
$^{1}$Science \& Technology Institute, Universities Space Research Association, 320 Sparkman Drive, Huntsville, AL 35805, USA\\
$^{2}$NASA Marshall Space Flight Center, Huntsville, AL 35812, USA\\
$^{3}$ University of Alabama in Huntsville, Huntsville, AL 35899, USA\\
$^{4}$Institute of Astrophysics, Foundation for Research and Technology-Hellas, GR-70013 Heraklion, Greece\\
$^{5}$ INAF Osservatorio Astronomico di Brera, Via E. Bianchi 46, 23807 Merate (LC), Italy\\
$^{6}$ Department of Physics, University of Crete, 70013, Heraklion, Greece \\
$^{7}$ INAF Osservatorio Astronomico di Roma, Via Frascati 33, 00078 Monte Porzio Catone, RM, Italy\\
}

\date{Accepted XXX. Received YYY; in original form ZZZ}

\pubyear{\the\year{}}

\begin{document}
\label{firstpage}
\pagerange{\pageref{firstpage}--\pageref{lastpage}}
\maketitle

\begin{abstract}
IXPE has enabled the X-ray polarizations of many blazars to be measured. We present the first population study for high synchrotron peaked blazars observed with IXPE using a uniform X-ray data analysis. We do not find any statistically significant correlations between the spectral properties of the blazars and their X-ray polarimetric properties. The lack of significant correlations may imply that there is little connection between the energy distribution of the X-ray emitting electrons and the uniformity of the magnetic field in the X-ray emitting regions of these blazars. These results will inform future theoretical work and potentially help narrow down the acceleration process of the synchrotron electrons.
\end{abstract}




\section{Introduction}\label{sec:intro}

Blazars are active galactic nuclei (AGN) with the jet pointed almost directly along {our line of sight \citep[e.g.][]{blandford2019}, and are particularly bright due to relativistic beaming effects. They emit across the entire electromagnetic spectrum \citep[e.g.,][]{hovatta2019}. The blazar spectral energy distribution (SED) is two-humped, with a lower energy peak attributed to synchrotron radiation, and a higher energy peak whose origin is not fully understood, with the strongest evidence pointing towards an inverse Compton origin \citep{liodakis2025}. A subset of blazars are known as high synchtrotron peaked blazars (HSPs). These are blazars where the synchrotron hump peaks above $10^{15}$ Hz. 

The Imaging X-ray Polarization Explorer (IXPE) has provided a new way to study blazars. IXPE is the first focusing X-ray observatory able to both conduct spectroscopy and polarimetry in the energy range of 2-8 keV. The observatory possesses three gas pixel detector units (DUs), which each create their own count (or I) spectra, as well as their own linear polarization (or Q and U) spectra. These spectra can be jointly fit together with ancillary X-ray observations in order to perform spectropolarimetry: the simultaneous measurement of spectral and polarimetric properties, with the ability to distinguish the contributions of individual spectral components to the polarization. Comparisons between spectral and polarimetric properties could be a new way to constrain theoretical models of particle acceleration in blazar jets.

Several HSPs have been observed by the IXPE team: 1ES 0229+200 \citep{ehlert2023}, 1ES 1959+650 \citep{errando2024,pacciani2025}, Mrk 421 \citep{DiGesu2022b,DiGesu2023,kim2024a,maksym2025}, Mrk 501 \citep{liodakis2022,chen2024,lisalda2026}, PG 1553+113 \citep{middei2023b,middei2026}, and PKS 2155-304 \citep{kouch2024}. All of them have synchrotron peaks high enough that IXPE's 2-8 keV band samples the synchrotron hump. 

The results of IXPE observations of HSPs indicate that the electrons that produce the synchrotron emission are heavily energy stratified \citep{marscher2024}. That is, the electrons producing the X-ray emission radiate in a smaller region close to the acceleration zone, whereas the optical emission arises from a larger region further downstream. The higher energy electrons capable of producing X-rays are rapidly slowed by synchrotron cooling, confining them to radiate X-rays in a region close to the particle acceleration site \citep{zhang2024}. Lower energy electrons make it further down the jet stream before losing energy, radiating in the lower energy bands like optical and radio in a larger and more distant region from the acceleration site \citep[e.g.,][]{tavecchio2018,tavecchio2021}. Higher X-ray polarizations than optical polarizations are seen in the majority of HSPs observed with IXPE \citep{marscher2024}. \citet{marscher1985} proposed that the explanation for this difference in polarization is that the electrons are accelerated in the first place by shocks. The magnetic field is more uniform closer to the shock where the high energy electrons emit X-rays, leading to higher polarization in the X-ray band as compared to the optical or radio bands \citep[e.g.,][]{tavecchio2020,tavecchio2021}. The higher polarization in the X-ray band could also be due to turbulence, in which case the lower number of turbulent cells occupied by the smaller X-ray emitting region does not dilute the polarization as much as the larger number of turbulent cells occupied by the optical-emitting region \citep[e.g.,][]{DiGesu2022a,marscher2024,sciaccaluga2025}. Shock acceleration also leads to a polarization angle parallel to the jet axis \citep{liodakis2022}, which is seen in all IXPE-observed HSPs within the uncertainties \citep{capecchiacci2025}. However, it is still not possible to rule out magnetic reconnection as the acceleration mechanism, as it can explain some of the HSPs' behavior, such as the polarization angle rotations more than 180 degrees seen in Mrk 421 \citep{DiGesu2023}. In either case, the key is that the X-ray emitting region is smaller and closer to the acceleration site than the optical-emitting region.

IXPE results on the HSPs cover a large range of polarization degrees (PDs), ranging from less than 5\% to almost 31\% in PKS 2155-304 \citep{kouch2024,marscher2024}. The ratio of X-ray to optical polarizations, $\Pi_{x}/\Pi_{o}$, also varies widely, from close to one to greater than seven \citep[]{marscher2024,capecchiacci2025}. It is an open question as to why this is the case. We desire to connect this diversity of X-ray PD behavior to the better established variability of other physical variables, like the shape of the X-ray spectrum, the SED, and the luminosity.

Here we perform the first population study of the X-ray spectropolarimetric properties of HSP blazars observed by IXPE. Our main goal is to investigate the potential relationships between the spectral properties (which are primarily determined by the electron energy distribution) and the polarization properties (which are primarily determined by the direction and orderliness of the magnetic field). Our paper is structured as followed: in Section \ref{sec:obs} we discuss the observations used in this paper. In Section \ref{sec:specpol} we describe the process of analyzing the data. In Section \ref{sec:results} we discuss the outcomes of the data analysis, and in Section \ref{sec:discussion} we state what the implications of our results might be for the current understanding of the particle acceleration process in HSPs. In all cases error bars on our measurements are 68\% or 90\% confidence.
\section{Observations and Data Reduction}\label{sec:obs}

We started with all HSP IXPE observations that had published papers from the IXPE collaboration at the time of writing. We then eliminated all observations from the starting list that had no X-ray polarization detected at 99\% confidence. For ancillary X-ray data we preferentially used XMM-Newton or NuSTAR observations due to their large bandpass and relatively high spectral resolution. Swift-XRT was used only when the former two were not available. We only used data that were taken simultaneously with the IXPE observations. A summary of the datasets used in this study can be seen in Table \ref{tab:xrayobs}. All count (or I) spectra were grouped to have at least 30 counts per bin so that $\chi^{2}$ statistics could be used. 

\subsection{IXPE}
Background rejection was performed on all IXPE observations following the procedure of \citet{DiMarco2023}.  The IXPE spectra were extracted using circular source regions 60\arcsec{} in radius and annular background regions with inner radius 2\arcmin{} and outer radius 5\arcmin{}. NEFF weighting was used. The I, Q, and U spectra were extracted using the PyXSPEC Python wrapper for XSPEC 12.14.1. The procedure used was the same as used in the IXPE Quickstart Data Analysis Guide v2.2. The IXPE observations of 1ES 0229+200 (obsID: 01006499) and PG 1553+113 (obsID: 02004999) consisted of two separate segments, while Mrk 421 obsID 02008199 consisted of four separate segments. We split these observations into their component segments using XSELECT and extracted spectra separately from each one. The second segment of PG 1553+113 did not have any detected X-ray polarization at 99\% confidence so it was eliminated from the sample.
Mrk 421 obsIDs 01003801 and 01003901 have non-detected X-ray polarization when using the time-averaged spectra; however it is known that the polarization angle (PA) rotated over the course of these observations by 180\degree or more \citep{DiGesu2023}. The non-detection in the time-averaged spectra is therefore an artifact of averaging out opposing PA directions. To account for the variation of the PA, we split each observation into 16 equal-length time bins, such that each time bin spanned $\mathrm{\sim6\;ks}$ or $\mathrm{\sim1.7\;hours}$, similar to the binning scheme used in \citet{DiGesu2023}. Separate IXPE spectra were then extracted from each time bin.

In all cases the Q and U spectra were left ungrouped. 

\subsection{XMM-Newton}
The XMM-Newton data were reduced using XMM-SAS v21.00. Only the PN camera data were used. PG 1553+113, 1ES 0229+200, and 1ES 1959+650 obsID 090211501 were observed in imaging modes. Circular source regions 20\arcsec{} and 30\arcsec{} in radius were used for PG 1553+113 and 1ES 0229+200, respectively, while the background regions were circular with radii 60\arcsec{} and 80\arcsec{}. 1ES 1959+650 obsID 090211501 used a circular background region 25\arcsec{} in radius. The background regions were used to check for background flares and none were observed. The XMM-SAS function \texttt{epatplot} was used to check for pileup in these sources, and pileup was found in 1ES 1959+650 obsID 090211501, so the source region for extraction was an annulus with inner radius 10\arcsec{} and outer radius 20\arcsec{}.

PKS 2155-304, the other observations of 1ES 1959+650, Mrk 501, and Mrk 421 were all observed in timing mode. Background regions with the x-coordinates $3<\mathrm{RAWX}<5$ were used for all timing mode observations. Source regions were centered on the bore and drawn to take in the majority of the flux. Mrk 421's timing mode observations were universally piled-up, so the centermost pixels centered on the bore were excised from the source region until \texttt{epatplot} showed pileup was negligible. 

We used the range 1.0-10.0 keV for fitting for most observations. We tested including 0.5-1.0 keV as well, and the results were consistent within 99\% confidence, except for 1ES 1959+650 XMM obsIDs 0902110801 and 0902111501, and Mrk 421 XMM obsID 0902112401. 1ES 1959+650 obsID 0902110801 was substantially piled up below 1 keV even after correcting the spectral extraction regions for pileup, so we ignored energies below 1 keV for it. 1ES 1959+650 obsID 090211501 and Mrk 421 obsID 090211501 had negligible pileup after correction, so we used the 0.5-10.0 keV range for fitting these XMM observations.

\subsection{NuSTAR}
NuSTAR data were analyzed using HEASOFT v6.34, using \texttt{nupipeline} to reduce the data and \texttt{nuproducts} to extract the spectra. Source spectra were extracted from circular regions 60\arcsec{} in radius, while background spectra were extracted from circular regions 100\arcsec{} in radius. We used energies 3.0 - 50.0 keV in the spectral fitting. We tested restricting the uppermost energy to where the source became fainter than background, and the results were within 99\% confidence uncertainties, so we kept the uppermost noticed energy at 50.0 keV for all sources.

\subsection{Swift-XRT}
Swift-XRT data were reduced using HEASOFT v6.34, using 
\texttt{xtrtpipeline} to reduce the data and XSELECT to extract the spectra. The photon-counting mode data were universally piled-up, so we preferentially used data taken in timing mode to avoid pileup. The timing mode spectra were extracted using 60\arcsec{} radius source and background regions. For 1ES 0229+200 obsID 0096571010, the source did not show up clearly in the timing mode observations, so we used photon counting mode. We extracted an annular source region for this observation using the procedure in Table 2 of \citet{middei2022} to minimize pileup. Energies from 1.0-10.0 keV were used in the spectral fitting for most sources. We tested including energies 0.5-1.0 keV in the spectral fitting, and the results were consistent within the 99\% confidence interval uncertainties with the exception of 1ES 1959+650 Swift obsIDs 00097164002, 00013906083, 00013906084, 00097164003, 00097164004, and 00097164005. For these Swift observations we used the 0.5-10.0 keV range for fitting.

\begin{table*}
\caption{Summary of X-ray Observations, and the references for the X-ray, optical, infrared, and radio data used.}
\begin{threeparttable}
\centering
\setlength{\tabcolsep}{3pt}
\begin{tabular}{lccllll}
\hline
Blazar & 
IXPE Obsid &
IXPE Start Date &
Swift &
XMM &
NuSTAR &
Ref\\
\hline
1ES 0229+200 &01006499	&2023-01-15	& {} & 0902110201 & {} & \citet{ehlert2023}\\
{} & {''} & 2023-01-27 & 00096571010 & {} & {} & ''\\
1ES 1959+650 & 01006201 &2022-05-03  & {} & 0902110801 & {} & \citet{errando2024}\\
{} & 02004801 & 2022-10-28 & {} & 0902111501 & 60801022002 & \citet{pacciani2025}\\
{} & 02250801 & 2023-08-14 & 00097164002, 00013906083, 00013906084& {} & {} &  ''\\
{} & {''} & {''} & 00097164003, 00097164004, 00097164005 & {} & {} & ''\\
Mrk 421 & 01003701 &2022-05-04 & {}  & 0902112401 & 60701031002 & \citet{DiGesu2022b}\\
{} & 01003801\tnote{a} & 2022-06-04 & 00031540053, 00031540054, 00031540055 & {} & {} & \citet{DiGesu2023} \\
{} &01003901\tnote{a} &2022-06-07 &00089326001, 00096557011  & {} & {} & ''\\
{} & 02004401 & 2022-12-06 & {} & 0902111701  & {} & \citet{kim2024a}\\
{} & 02008199 & 2023-12-06 & {} & 0902112401, 0902112501  & 60902024002 & \citet{maksym2025}\\
{} & {''} & 2023-12-11 & {} & 0920900201, 0920901301  & 60902024004 & ''\\
{} & {''} & 2023-12-16 & {} & {} & 60902024006 & ''\\
{} & {''} & 2023-12-20 & {} & {} & 60902024008 & ''\\
Mrk 501 & 01004501 & 2022-03-07 & {} & {} & 60701032002 & \citet{chen2024}\\
{} & 01004601 & 2022-03-27 & {} & {} & 60702062004 & ''\\
{}&02004701 &2023-04-16 &00015411049, 00015411050  & {} & {} & ''\\
PG 1553+113 &02004999\tnote{b}	&2023-02-06	& {} & 0902112101 & {} & \citet{middei2023b}\\
PKS 2155-304 & 02005601	& 2023-10-27 & {} & 0902110901 & {} & \citet{kouch2024}\\
\hline
\end{tabular}
\begin{tablenotes}
\item[a]{Rotation observation. Divided into 16 bins.}
\item[b]{Only first epoch of observation is used}
\end{tablenotes}
\end{threeparttable}
\label{tab:xrayobs}
\end{table*}

\section{Spectropolarimetric Fitting}\label{sec:specpol}

Spectropolarimetric fitting is the process of simultaneously fitting the I, Q, and U spectra of a source in order to determine how both the flux and the polarization properties vary with energy. Spectral models are multiplied with polarization models, and then fit to the I, Q, and U spectra in a similar manner to how ordinary count spectra are fitted. 

Spectropolarimetric fitting of the data was done in XSPEC v12.15.1 \citep{arnaud96} using the default cosmology $\mathrm{H_{0}=70\;km/s/Mpc}$, $q_{0}=0$, and $\Lambda_{0}=0.73$. Each IXPE pointing/segment was fit separately with ancillary X-ray data simultaneous to it. 

We perform a uniform analysis, assuming the same spectral model for all the blazars, which has not been done before. We fit all of the I spectra using a \texttt{constant*tbabs*zlopgar} model, where the \texttt{constant} represents the cross-normalization constants between the different telescopes and DUs, \texttt{tbabs}
\citep{wilms2000} represents Galactic absorption, and \texttt{zlogpar} \citep{massaro2004} is a log-parabola model that takes into account redshift. The specific equation for \texttt{zlogpar} is expressed in the form 
\begin{equation}
N(E)=K[E(1+z)/E_{p}]^{(\alpha-\beta log(E(1+z)/E_{p}))}
\end{equation}
In this equation, $E_{p}$ is the pivot energy, $\alpha$ is the spectral slope at the pivot energy and $\beta$ is an indicator of spectral curvature at the synchrotron peak.  $K$ is a normalization constant with units of $\mathrm{photons\;cm^{-2}\;s^{-1}\;keV^{-1}}$. We fixed the pivot energy to 3 keV for all the blazars. We recorded the best fit $\alpha$, $\beta$, and normalization of the I spectra, and used them as  the initial parameters for the spectropolarimetric fits, which used the model \texttt{constant*tbabs*polconst*zlogpar}.  \texttt{polconst} is a polarization model that assumes a constant PD and PA with energy. We then refit the spectropolarimetric data  to obtain the final spectral and polarization parameters. $\alpha$, $\beta$, and $\Pi_x$ are tabulated in the third, fourth, and fifth columns of Table \ref{tab:specpol}.

\subsection{Time-Averaged Spectra}
In Mrk 501 IXPE obsIDs 01004501, the spectrum for DU2 deviated significantly from that of DU1 and DU3 around 2 keV and 8 keV, so for this observation we had to remove DU2 to get a good fit. The change in $\chi^2$/dof was from 2102/1903 to 1538/1492}. We had to allow the $\alpha$ and/or $\beta$ parameters to vary between the IXPE and ancillary spectra for several observations in order to get a good fit. These were 1ES 1959+650 obsID 02250801; Mrk 421 obsID 01003701 and the second and third segments of Mrk 421 obsID 02008199; Mrk 501 obsID 01004601; and PKS 21455-304 obsID 02005601. Allowing the $\alpha$ and $\beta$ parameters to vary is acceptable as the ancillary X-ray observations only covered part of the period over which IXPE observed the source, during which the spectral properties could have varied. A varying alpha was used in the IXPE discovery paper of PKS 2155-304 \citep{kouch2024}. We use the $\alpha$ and $\beta$ parameters of the IXPE spectrum in our further analysis for these sources. 

As a check on our results, we also performed fits on each of the blazar observations using only normal \texttt{logpar} like the previous IXPE discovery papers, and using the same pivot energies as had been previously used in the literature. The results of the spectropolarimetric fitting using these parameters were broadly consistent with those in the previous literature.

For comparison to the X-ray polarization, we pulled the average optical polarizations for each IXPE segment from the IXPE discovery papers on each blazar. We used the average R-band polarization measurements from each observation segment, as these were the only polarization measurements corrected for {\bf depolarization} from the host galaxy. 
Due to the fact that some of the HSPs have synchrotron peaks well below 1 keV, we calculated the synchrotron peak for each observation using not only the X-ray spectra but also simultaneous optical, infrared, and radio flux densities derived from the blazars' IXPE discovery papers (see the Ref column in Table \ref{tab:xrayobs}). The details of the data analysis for these flux densities and the telescopes used are described in the corresponding references. For the X-ray flux densities we used the X-ray spectral model, corrected for cross-calibration constants and absorption, and assumed the errors on the model points were the same relative scale as the errors on the original data points. We plotted the X-ray, optical, infrared, and radio data in $\nu$ vs $\nu L_\nu$ space and fit a generic parabolic model to it in Python using \texttt{numpy.polynomial.polynomial.polyfit}. We then used this model to find the vertex of the parabola, which became our value for the synchrotron peak of the SED. The fourth segment of Mrk 421 ObsID 02008199 possessed only X-ray data, so we computed the synchrotron peak using the equation for the synchrotron peak energy:
\begin{equation}
E_{sp}=E_{p}*10^{(2-\alpha)/(2\beta)}
\end{equation}
We used the $\alpha$ and $\beta$ values from the IXPE spectrum.

We were unable to calculate the synchrotron peak of 1ES 0229+200 as its peak was well above the IXPE band and no simultaneous NuSTAR data was available. For this blazar we instead used the synchrotron peak found by \citet{Costamante2018}.
The logarithm of each observation's measured synchrotron peak in Hz is tabulated in the eighth column of Table \ref{tab:specpol}.

From each observation's best fit to the X-ray spectrum, we measured the unabsorbed fluxes in the 2-4 keV and 4-8 keV bands using the \texttt{cflux} component in XSPEC, and the unabsorbed 2-10 keV luminosity using the \texttt{clumin} component in XSPEC. In both cases we corrected for cosmological redshift. We used the 2-4 and 4-8 keV fluxes to calculate the hardness ratio, $\mathrm{F_{4-8}/F_{2-4}}$. The dimensionless hardness ratio and logarithm of the luminosity in $\mathrm{erg\;s^{-1}}$ are tabulated in the ninth and tenth columns of Table \ref{tab:specpol}.

\subsection{Time-Resolved Mrk 421 Spectropolarimetry}
We fit each of the 16 spectra for obsIDs 01003801 and 01003901 using the same procedure as above. We used the Python wrapper for XSPEC, PyXSPEC, to loop over each bin and find the best fit for it. For each bin, we used IXPE spectra extracted solely from the duration of the bin, and jointly fit them with the Swift spectra from the entire observation. Several of these time bins had 99\% confidence intervals on $\Pi_x$ consistent with zero, and so were removed from the final sample.

The optical PD varied much more slowly than the X-ray PD, so we use the average optical PD of each entire observation in future analysis, which was $\mathrm{\Pi_o=4.35\pm0.1\%}$ for obsID 01003801 and $\mathrm{\Pi_o=5.4\pm0.3\%}$ for obsID 01003901.

$\mathrm{log(\nu_{sp}/Hz)}$, $\mathrm{F_{4-8}/F_{2-4}}$, and $\mathrm{log(L_{2-10})}$ were calculated for each time bin in the same manner as they were for the time-averaged spectra. The results are tabulated in Table \ref{tab:Mrk421obs2} and Table \ref{tab:Mrk421obs3}.

\section{Results}\label{sec:results}

The results of the spectropolarimetric fits and calculations can be seen for the time-averaged observations in Table \ref{tab:specpol}. 
PKS 2155-304 had the highest X-ray polarization in the entire sample with {\bf $\Pi_x=$} 23\%. This is consistent with previous work showing that PKS 2155-304 had the highest X-ray polarization found in any blazar observed by IXPE \citep{kouch2024}. The $\alpha$ parameter varied mostly between 2 and 2.8, with the exception of the first observation of 1ES 1959+650 and the second observation of Mrk 501 which had an unusually low values of $\alpha$ at 1.89 and 1.73, respectively.  The $\beta$ parameter varied between 0.02 for the third observation of Mrk 501 and 0.91 for the first observation of Mrk 421. 

\begin{table*}
\caption{Time-Averaged Sample Spectropolarimetry Results}
\begin{threeparttable}
\centering
\setlength{\tabcolsep}{3pt}
\begin{tabular}{lcccccccccc}
\hline
Blazar & 
IXPE Obsid &
Energies (keV) &
$\alpha$ &
$\beta$ &
$\Pi_{x}$ (\%) &
$\Pi_{o}$ (\%) &
$\mathrm{log(\nu_{sp}/Hz)}$ &
$\mathrm{F_{4-8}/F_{2-4}}$ &
$\mathrm{log(L_{2-10})}$ &
$\mathrm{\chi^{2}/d.o.f.}$\\
\hline
1ES 0229+200 &01006499 & 1.0-10.0	& $1.91\pm0.02$	& $0.27^{+0.34}_{-0.32}$\tnote{a} & $11\pm6$ &  $2.42\pm0.72$ & $18.34\pm0.01$\tnote{b} & $1.05\pm0.05$ & {$44.72\pm0.02$} &{1330.11/1182}\\
{} & {"} & {1.0-10.0}	& {$2.04\pm0.07$} & {$0.57^{+0.36}_{-0.35}$} & {$13^{+6}_{-5}$} & {$3.2\pm0.7$}\tnote{c} & {"} & $0.87\pm0.03$ & $44.81\pm0.02$ & {1068.90/1099} \\
1ES 1959+650 &	01006201 & 1.0-10.0 & $1.89\pm0.03$ &  $0.51\pm0.19$ & $6\pm3$ & $4.49\pm{0.17}$ &  $17.02\pm0.01$ & $0.08\pm0.03$ & $45.01\pm0.01$ &{1270.80/1224}\\
{} & 02004801 & 0.5-50.0 & $2.37\pm0.01$ & $0.16\pm0.02$& $10\pm2$ & $4.54\pm0.7$ & $16.85\pm0.01$ & $0.76\pm0.01$ & $44.74\pm0.01$ &{1448.49/1381}\\
{} & 02250801 & 0.5-10.0 & $2.64\pm0.01$\tnote{a} & $0.29\pm0.06$\tnote{a} & $12\pm1$ & $5.48\pm0.47$ & $16.52\pm0.01$ & $0.60\pm0.01$ & $45.91\pm0.1$ & {3114.64/2940}\\
Mrk 421 & 01003701 & 0.5-50.0 & $2.38\pm0.03$\tnote{a} & $0.91\pm0.18$\tnote{a} & $14^{+3}_{-2}$ & $2.9\pm0.1$ & $16.38\pm0.03$ & $0.66\pm0.01$ & $44.41\pm0.01$ & {1990.81/1845}\\
{} & 02004401 & 1.0-10.0 & $2.73\pm0.01$ & $0.18\pm0.03$ & $13\pm2$ & $4.6\pm1.3$ &  $16.63\pm0.01$ & $0.59\pm0.01$ & $44.73\pm0.01$ & {1350.40/1338}\\
{} & 02008199  & 1.0-50.0 & $2.70\pm0.01$ & $0.17\pm0.01$ & $10\pm1$ & $4.85\pm0.10$ & $15.78\pm0.01$ & $0.62\pm0.01$ & $44.73\pm0.01$ & {2124.18/2131}\\
{} & {''} & 1.0-50.0 & $2.63\pm0.01$\tnote{a} & $0.23\pm0.08$\tnote{a} & $16\pm1$ & $3.78\pm0.09$ & $15.99\pm0.01$ & $0.62\pm0.01$ & $44.93\pm0.01$ &{2213.64/2128}\\
{} & {''} & 2.0-50.0 & $2.75\pm0.01$ & $0.26\pm0.02$ & $9\pm1$ & $5.29\pm0.38$ & $15.74\pm0.01$ & $0.58\pm0.01$ & $44.91\pm0.01$ & {2204.28/1969}\\
{} & {''} & 2.0-50.0 & $2.78\pm0.02$\tnote{a} & $0.23\pm0.17$\tnote{a} & $6\pm2$ & $5.28\pm0.26$ & $16.20\pm0.08$\tnote{d} & $0.53\pm0.01$ & $44.88\pm0.01$ & {2073.23/1954}\\
Mrk 501 & 01004501\tnote{e} & {2.0-50.0}	& {$2.10\pm0.03$} & {$0.34\pm0.03$} & {$9\pm3$} & {$6.6\pm0.4$} & {$16.82\pm0.01$} & $0.87\pm0.02$ & $44.50\pm0.01$ & {1538.41/1492}\\
{} & 01004601 & {2.0-50.0}	& {$1.73\pm0.02$}\tnote{a} & {$0.02^{+0.15}_{-0.14}$} & {$9\pm2$} & {$4.7\pm0.3$} & {$17.39\pm0.01$} & $1.20\pm0.02$ & $44.85\pm0.01$ &{2093.75/2056} \\
{}&02004701 & 1.0-10.0 & $2.32\pm0.03$ & $0.35\pm0.16$ & $16\pm3$ & $5.9\pm1.5$ & $16.54\pm0.23$  & $0.74\pm0.01$& $44.25\pm0.01$ & {1134.64/1225}\\
PG 1553+113  & 02004999\tnote{f}	& 1.0-10.0 & $2.46\pm0.01$	&  $0.08\pm0.04$ & $8\pm5$ & $2.2\pm0.4$ & $15.42\pm0.01$ & $0.72\pm0.01$ & $46.54\pm0.01$ & {1226.46/1216}\\
PKS 2155-304 & 02005601	&  1.0-10.0 & $2.69\pm{0.03}$\tnote{a} & $0.48\pm0.17$\tnote{a} & $23\pm2$ & $3.9\pm0.22$ & $15.99\pm0.01$ & $0.56\pm0.01$ & $44.88\pm0.08$ & {1429.00/1330}\\
\hline
\end{tabular}
\begin{tablenotes}
\item[a]{Value is from the IXPE spectrum only.}
\item[b]{From \citet{Costamante2018}}
\item[c]{Not exactly simultaneous (within 3 days)}
\item[d]{Calculated using only X-ray spectrum}
\item[e]{DU2 omitted}
\item[f]{Only the first epoch used}
\end{tablenotes}
\end{threeparttable}

\label{tab:specpol}
\vspace{0.5cm}
$\alpha$, $\beta$, and $\Pi_x$ error bars are 90\% confidence intervals, while $\Pi_o$, $\mathrm{log(\nu_{sp}/Hz)}$, $\mathrm{F_{4-8}/F_{2-4}}$, $\mathrm{F_{4-8}/F_{2-4}}$, and $\mathrm{log(L_{2-10})}$ error bars are 68\% confidence intervals. Fluxes and luminosities are corrected for cosmological redshift and Galactic absorption.
\end{table*}

\begin{table*}
\caption{Mrk 421 01003801 Time Resolved Spectropolarimetry Results}
\centering
\setlength{\tabcolsep}{3pt}
\begin{tabular}{lcccclllllll}
\hline
Bin \# &
Bin Start (MJD) &
Bin stop (MJD) &
$\alpha$ &
$\beta$ &
$\Pi_{x}$ (\%) &
$\Pi_o$ (\%) & 
$\mathrm{log(\nu_{sp}/Hz)}$ &
$\mathrm{F_{4-8}/F_{2-4}}$ &
$\mathrm{log(L_{2-10})}$ &
$\mathrm{\chi^{2}/d.o.f.}$\\
\hline
1 & 59734.47 & 59734.60	& $2.55\pm0.03$	& $0.18\pm{0.04}$ & $15\pm7$ & $4.35\pm0.1$ & $15.33\pm0.01$ & $0.66\pm0.02$ & {$43.53\pm0.01$} &{1533.29/1669}\\
2 & {59734.60} & {59734.72} & {$2.56\pm0.03$} & {$0.19\pm0.04$} & {$12\pm7$} & " & {$15.29\pm0.01$} & $0.66\pm0.02$ & $43.49\pm0.01$ & {1629.18/1663} \\
4 & {59734.84} & {59734.97} & {$2.58\pm0.03$} & {$0.21\pm0.04$} & {$11\pm8$} & " & {$15.13\pm0.01$} & $0.65\pm0.02$ & $43.36\pm0.02$ & {1434.44/1647} \\
14 & {59736.08} & {59736.21} & {$2.54\pm0.03$} & {$0.17\pm0.04$} & {$12\pm6$} & " & {$15.28\pm0.01$} & $0.69\pm0.01$ & $43.48\pm0.01$ & {1606.37/1665} \\
15 & {59736.21} & {59736.33} & {$2.51\pm0.03$} & {$0.12\pm0.04$} & {$11\pm6$} & " & {$15.49\pm0.01$} & $0.69\pm0.01$ & $43.72\pm0.01$ & {1749.07/1698} \\
16 &{59736.33} & {59736.46} & {$2.53\pm0.03$} & {$0.16\pm0.04$} & {$12\pm6$} & " & {$15.43\pm0.01$} & $0.68\pm0.01$ & $43.65\pm0.01$ & {1664.67/1686} \\
\hline
\end{tabular}
\label{tab:Mrk421obs2}
\\Only time bins with detections are shown. $\alpha$, $\beta$, and $\Pi_x$ error bars are 90\% confidence intervals, while $\Pi_o$ , $\mathrm{log(\nu_{sp}/Hz)}$, $\mathrm{F_{4-8}/F_{2-4}}$, $\mathrm{F_{4-8}/F_{2-4}}$, and $\mathrm{log(L_{2-10})}$ error bars are 68\% confidence intervals. Fluxes and luminosities are corrected for cosmological redshift and Galactic absorption.
\end{table*}

\begin{table*}
\caption{Mrk 421 01003901 Time Resolved Spectropolarimetry Results}
\centering
\setlength{\tabcolsep}{3pt}
\begin{tabular}{lcccclllllll}
\hline
Bin \# &
Bin Start (MJD) &
Bin stop (MJD) &
$\alpha$ &
$\beta$ &
$\Pi_{x}$ (\%) &
$\Pi_o$ (\%) & 
$\mathrm{log(\nu_{sp}/Hz)}$ &
$\mathrm{F_{4-8}/F_{2-4}}$ &
$\mathrm{log(L_{2-10})}$ &
$\mathrm{\chi^{2}/d.o.f.}$\\
\hline
3 & {59737.63} & {59737.76} & {$2.33\pm0.02$} & {$0.34\pm0.03$} & {$10\pm5$} & $5.4\pm0.3$ & {$16.05\pm0.02$} & $0.66\pm0.01$ & $43.81\pm0.01$ & {1604.78/1653} \\
5 & {59737.88} & {59738.01} & {$2.34\pm0.02$} & {$0.35\pm0.03$} & {$8\pm4$} & " & {$16.04\pm0.02$} & $0.65\pm0.01$ & $43.80\pm0.02$ & {1632.85/1653} \\
6 & {59738.01} & {59738.14} & {$2.33\pm0.02$} & {$0.34\pm0.03$} & {$14^{+6}_{-5}$} & " & {$15.92\pm0.02$} & $0.65\pm0.01$ & $43.70\pm0.01$ & {1570.54/1642} \\
7 & {59738.14} & {59738.26} & {$2.35\pm0.02$} & {$0.36\pm0.03$} & {$10\pm5$} & " & {$16.02\pm0.02$} & $0.66\pm0.01$ & $43.78\pm0.01$ & {1618.89/1651} \\
9 & {59738.39} & {59738.52} & {$2.35\pm0.02$} & {$0.36\pm0.03$} & {$10\pm5$} & " & {$16.03\pm0.02$} & $0.68\pm0.01$ & $43.80\pm0.01$ & {1608.47/1656} \\
10 & {59738.52} & {59738.64} & {$2.35\pm0.02$} & {$0.36\pm0.03$} & {$8^{+6}_{-5}$} & " & {$15.90\pm0.02$} & $0.66\pm0.01$ & $43.68\pm0.01$ & {1576.81/1635} \\
13 & {59738.90} & {59739.02} & {$2.34\pm0.02$} & {$0.34\pm0.03$} & {$11\pm6$} & " & {$15.83\pm0.02$} & $0.66\pm0.01$ & $43.63\pm0.01$ & {1518.12/1626} \\
16 & {59739.28} & {59739.40} & {$2.35\pm0.02$} & {$0.36\pm0.03$} & {$8^{+6}_{-5}$} & " & {$15.90\pm0.02$} & $0.68\pm0.01$ & $43.68\pm0.01$ & {1574.61/1631} \\
\hline
\end{tabular}
\\Only time bins with detections are shown. $\alpha$, $\beta$, and $\Pi_x$ error bars are 90\% confidence intervals, while $\Pi_o$ , $\mathrm{log(\nu_{sp}/Hz)}$, $\mathrm{F_{4-8}/F_{2-4}}$, $\mathrm{F_{4-8}/F_{2-4}}$, and $\mathrm{log(L_{2-10})}$ error bars are 68\% confidence intervals. Fluxes and luminosities are corrected for cosmological redshift and Galactic absorption.
\label{tab:Mrk421obs3}
\end{table*}

\begin{table*}
\begin{threeparttable}
\caption{Potential Correlations and their Statistical Significance for the Entire Blazar Sample.}
\begin{tabular}{llccc}
\hline
Independent variable & 
Dependent Variable &
$\mathrm{r_{s}}$\tnote{a} &
p-value\tnote{b} &
MC Results\tnote{c}\\
\hline
$\alpha$ & $\Pi_x$ & 0.23& 0.22 & 4.04\%\\
$\beta$ & $\Pi_x$ & 0.01 & 0.97 & 1.30\%\\
$\mathrm{log(\nu_{sp}/Hz)}$ & $\Pi_x$ & -0.16& 0.40 & 3.12\%\\
$\mathrm{F_{4-8}/F_{2-4}}$ & $\Pi_x$ & -0.05 & 0.40 &0.04\%\\
$\mathrm{log(L_{2-10})}$ & $\Pi_x$ & -0.23 & 0.23 & 2.91\%\\
$\Pi_o$ & $\Pi_x$ & -0.33 & 0.08 & 15.75\%\\
$\alpha$ & $\Pi_x/\Pi_{o}$  & 0.21 & 0.26 & 2.80\%\\
$\beta$ & $\Pi_x/\Pi_{o}$ & -0.15 & 0.43 & 2.29\%\\
$\mathrm{log(\nu_{sp}/Hz)}$ & $\Pi_x/\Pi_{o}$ & -0.13 & 0.43 & 1.70\%\\
$\mathrm{F_{4-8}/F_{2-4}}$ & $\Pi_x/\Pi_{o}$ & 0.10 & 0.60 & 0.10\%\\
$\mathrm{log(L_{2-10})}$ & $\Pi_x/\Pi_{o}$ & -0.10 & 0.59 & 0.19\%\\
\hline
\end{tabular}
\begin{tablenotes}
\item[a]{Spearman rank coefficient. Does not take into account error bars}
\item[b]{p-value from Spearman rank correlation test. Does not take into account error bars.}
\item[c]{Percentage of time Spearman rank correlation test resulted in statistical significance during Monte Carlo simulations. Takes into account error bars.}
\end{tablenotes}
\label{tab:corr}
\end{threeparttable}
\end{table*}

\begin{table*}
\begin{threeparttable}
\caption{Potential Correlations and their Statistical Significance for Mrk 421 obsID 01003801}
\begin{tabular}{llcc}
\hline
Independent variable & 
$\mathrm{r_{s}}$ &
p-value &
MC Results\\
\hline
$\alpha$ &  0.82 & 0.05 & 7.98\%\\
$\beta$  & 0.82 & 0.05 & 7.89\%\\
$\mathrm{log(\nu_{sp}/Hz)}$  & -0.70& 0.12 & 6.96\%\\
$\mathrm{F_{4-8}/F_{2-4}}$ & -0.87 & 0.02 &7.17\%\\
$\mathrm{log(L_{2-10})}$ &  -0.70 & 0.12 & 6.93\%\\
\hline
\end{tabular}
\label{tab:corr_Mrk421obs2}
\end{threeparttable}
\\. Correlations with $\Pi_x/\Pi_o$ are omitted because $\Pi_o$ was assumed to be constant over the entire observation.
\end{table*}

\begin{table*}
\begin{threeparttable}
\caption{Potential Correlations and their Statistical Significance for Mrk 421 obsID 01003901}
\begin{tabular}{llcc}
\hline
Independent variable & 
$\mathrm{r_{s}}$ &
p-value &
MC Results\\
\hline
$\alpha$ &  -0.54& 0.17 & 6.73\%\\
$\beta$ &  -0.63 & 0.09 & 6.72\%\\
$\mathrm{log(\nu_{sp}/Hz)}$ & -0.10& 0.81 & 4.06\%\\
$\mathrm{F_{4-8}/F_{2-4}}$  & -0.24 & 0.56 & 4.89\%\\
$\mathrm{log(L_{2-10})}$  & -0.06 & 0.88 & 3.59\%\\
\hline
\end{tabular}
\label{tab:corr_Mrk421obs3}
\end{threeparttable}
\\Correlations with $\Pi_x/\Pi_o$ and $\Pi_o$ are omitted because $\Pi_o$ was assumed to be constant over the entire observation.
\end{table*}

\begin{table*}
\begin{threeparttable}
\caption{Potential Correlations and their Statistical Significance for Mrk 421 obsIDs 01003801 and 01003901 Combined}
\begin{tabular}{llcc}
\hline
Independent variable & 
$\mathrm{r_{s}}$ &
p-value &
MC Results\\
\hline
$\alpha$  & 0.55 & 0.04 & 15.53\%\\
$\beta$ & -0.69 & 0.01 & 17.67\%\\
$\mathrm{log(\nu_{sp}/Hz)}$ &  -0.71 & 0.01 & 23.83\%\\
$\mathrm{F_{4-8}/F_{2-4}}$ & -0.26 & 0.36& 4.82\%\\
$\mathrm{log(L_{2-10})}$  & -0.66 & 0.01 & 19.97\%\\
\hline
\end{tabular}
\label{tab:corr_Mrk421all}
\end{threeparttable}
\\Correlations with $\Pi_x/\Pi_o$ and $\Pi_o$ are omitted because $\Pi_o$ was assumed to be constant over the entire observation.
\end{table*}

\begin{figure*}
\centering
\begin{minipage}{.5\textwidth}
\includegraphics[scale=0.50]{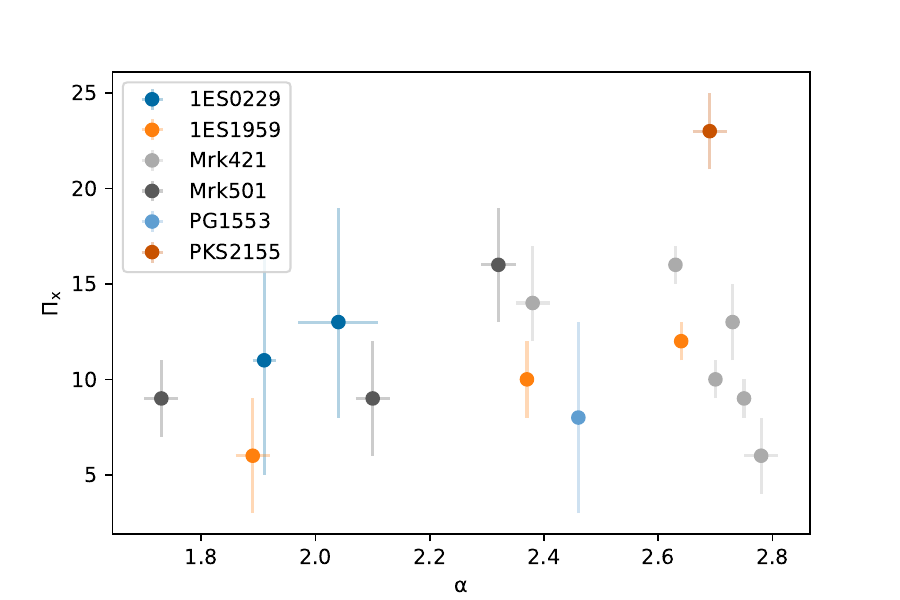}
\end{minipage}%
\begin{minipage}{.5\textwidth}
\includegraphics[scale=0.50]{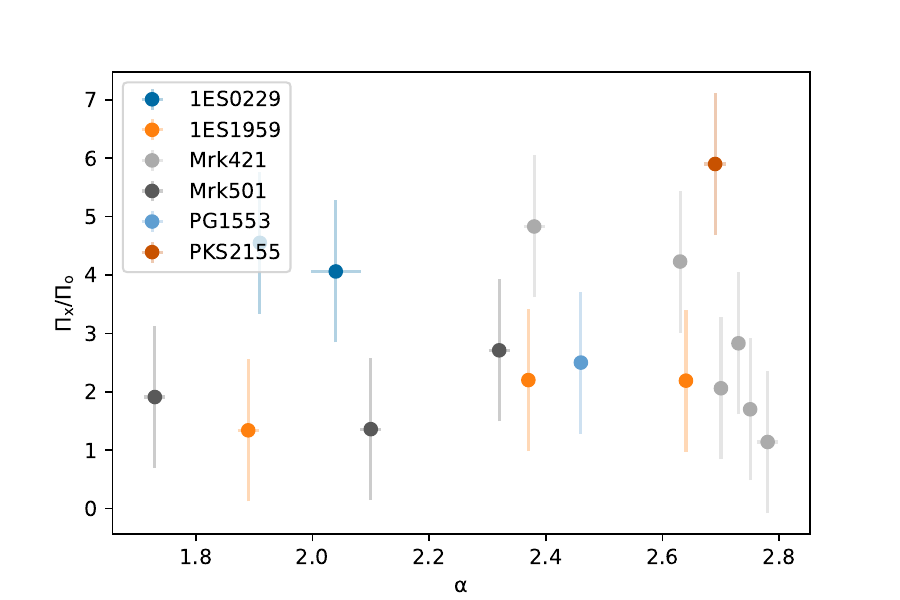}
\end{minipage}
\caption{Plot of the \texttt{zlogpar} parameter $\alpha$ vs the X-ray polarization degree (left) and X-ray to optical polarization ratio (right) for the time-averaged observations. Error bars on the left plot are 90\% confidence intervals, error bars on the right plot are 68\% confidence intervals. \label{fig:alpha}}
\end{figure*}

\begin{figure*}
\centering
\begin{minipage}{.5\textwidth}
\includegraphics[scale=0.5]{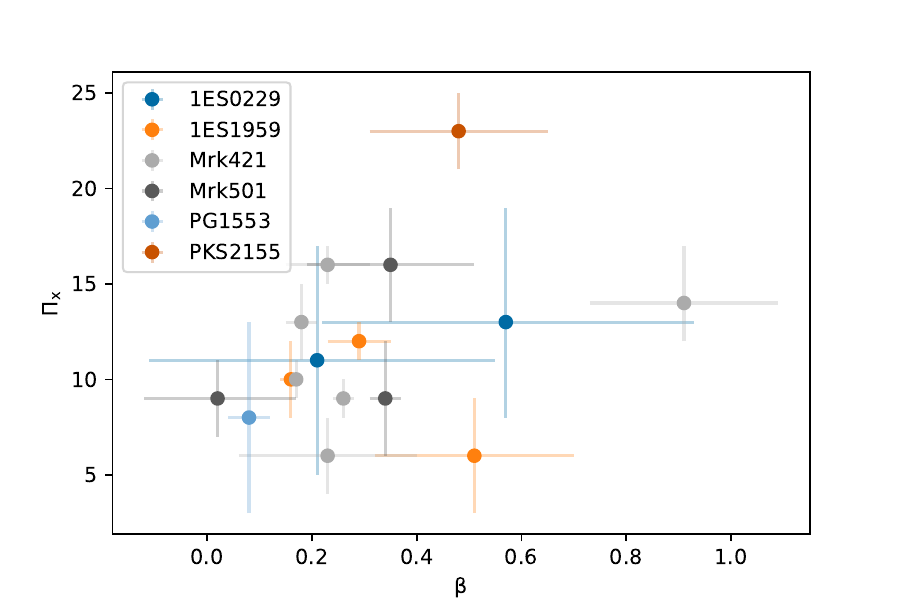}
\end{minipage}%
\begin{minipage}{.5\textwidth}
\includegraphics[scale=0.5]{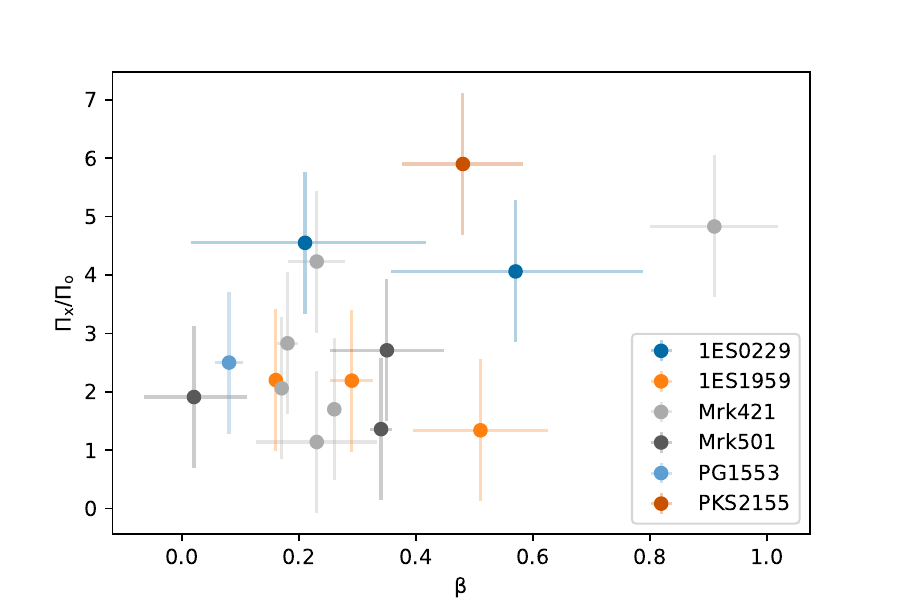}
\end{minipage}
\caption{Plot of the \texttt{zlogpar} parameter $\beta$ vs the X-ray polarization degree (left) and X-ray to optical polarization ratio (right) for the time averaged observations. Error bars on the left plot are 90\% confidence intervals, error bars on the right plot are 68\% confidence intervals.  \label{fig:beta}}
\end{figure*}

\begin{figure*}
\centering
\begin{minipage}{.5\textwidth}
\includegraphics[scale=0.5]{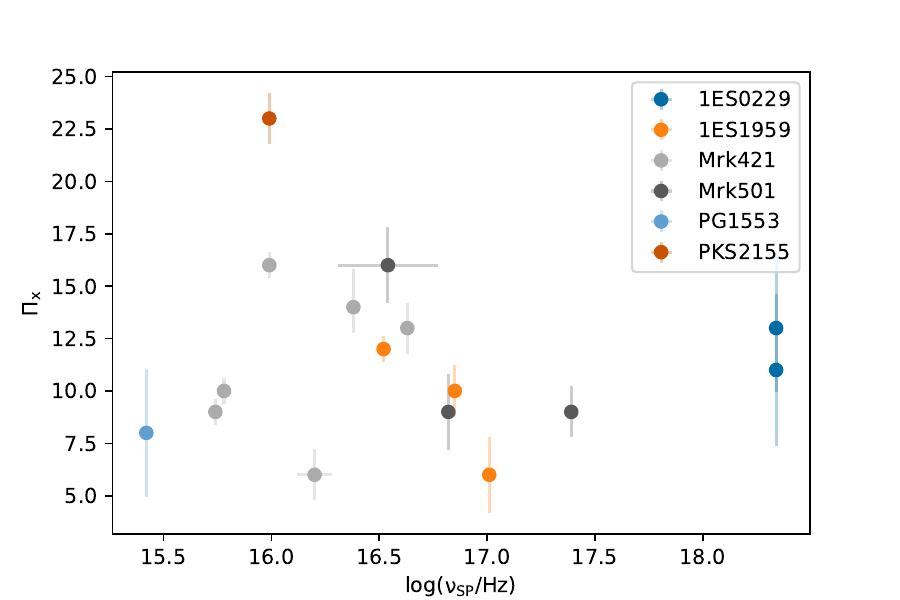}
\end{minipage}%
\begin{minipage}{.5\textwidth}
\includegraphics[scale=0.5]{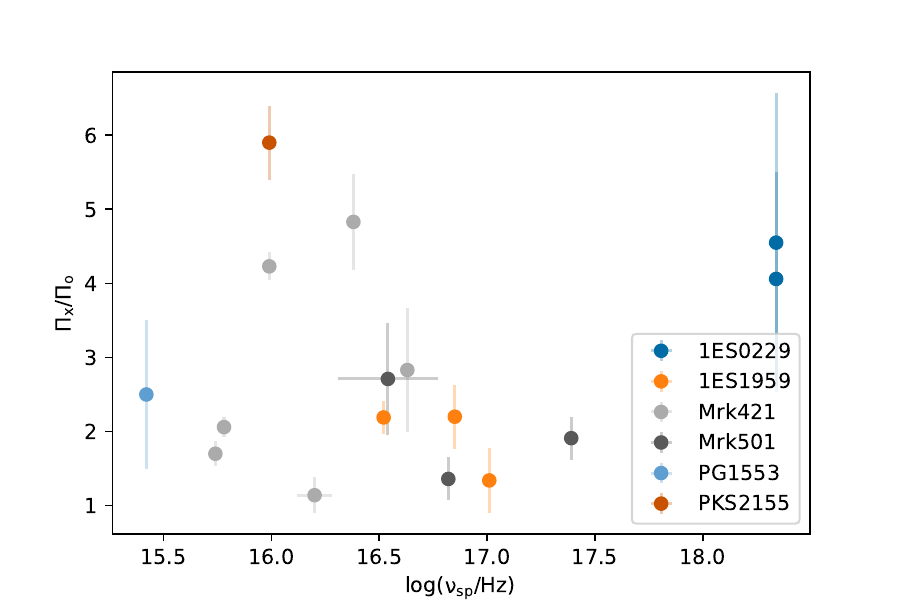}
\end{minipage}
\caption{{\bf Plot of the logarithm of the synchrotron peak frequency} vs the X-ray polarization degree (left) and the X-ray to optical polarization ratio (right) for the time-averaged observations. Error bars are 68\% confidence intervals. \label{fig:sp}}
\end{figure*}

\begin{figure*}
\centering
\begin{minipage}{.5\textwidth}
\includegraphics[scale=0.5]{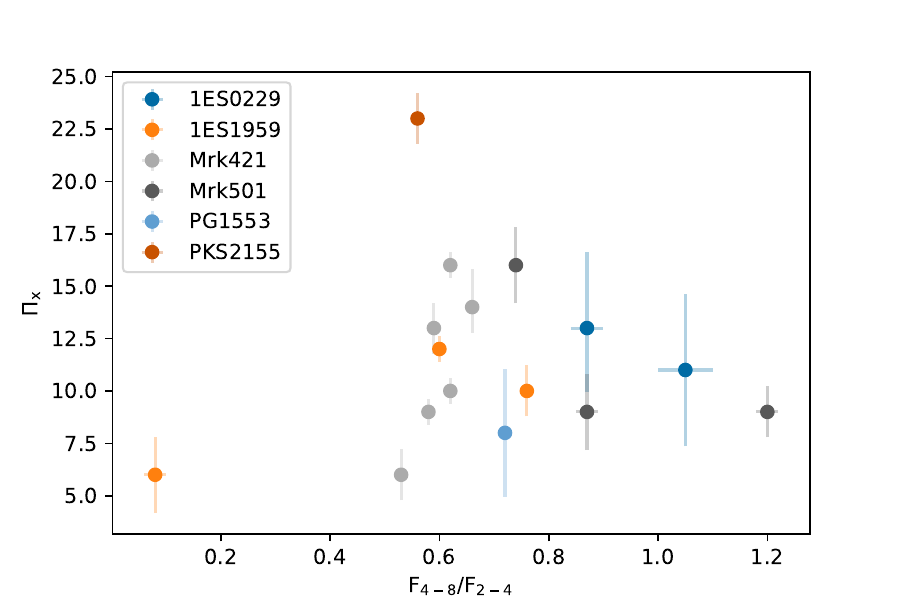}
\end{minipage}%
\begin{minipage}{.5\textwidth}
\includegraphics[scale=0.5]{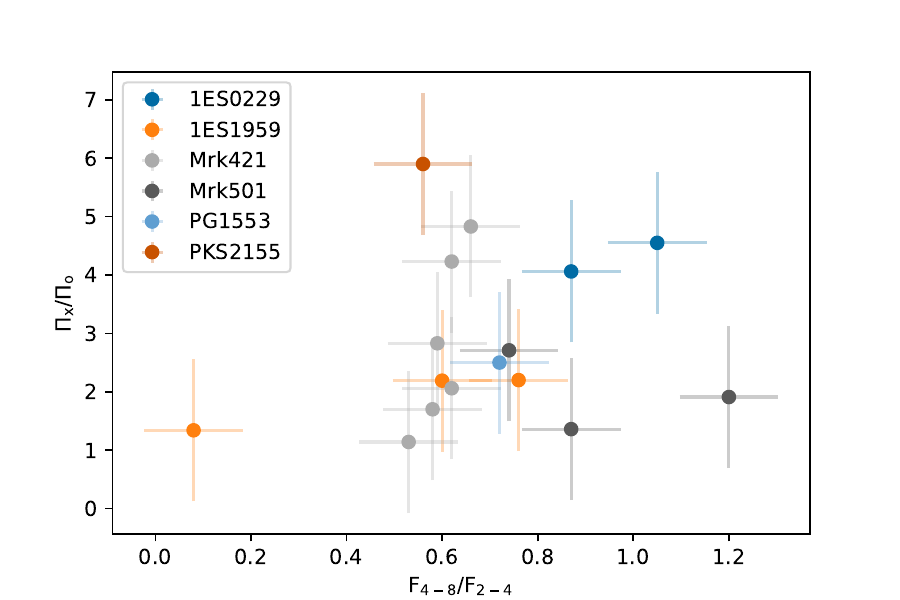}
\end{minipage}
\caption{Plot of the hardness ratio vs the X-ray polarization degree (left) and the X-ray to optical polarization ratio (right) for the time-averaged spectra. Error bars are 68\% confidence intervals.  \label{fig:HR}}
\end{figure*}

\begin{figure*}
\centering
\begin{minipage}{.5\textwidth}
\includegraphics[scale=0.5]{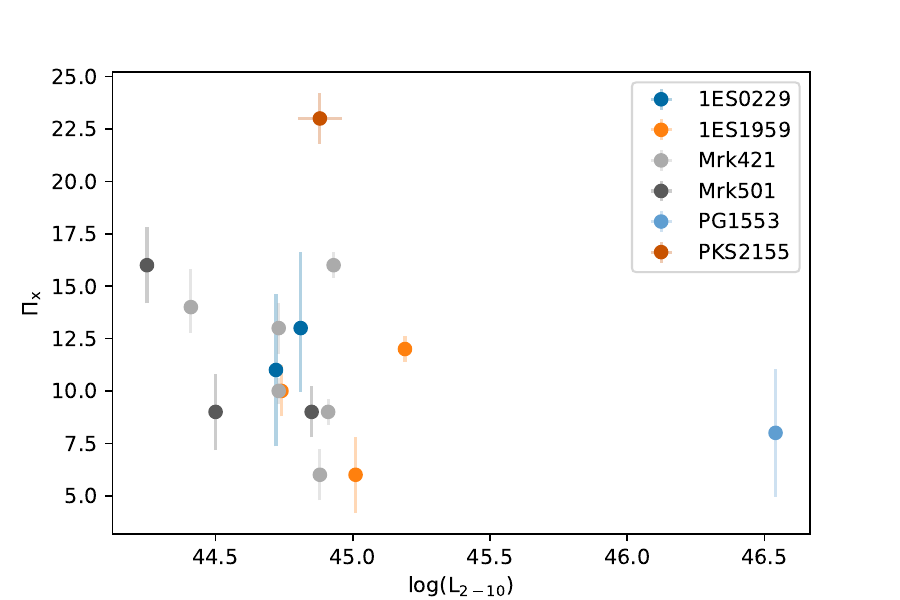}
\end{minipage}%
\begin{minipage}{.5\textwidth}
\includegraphics[scale=0.5]{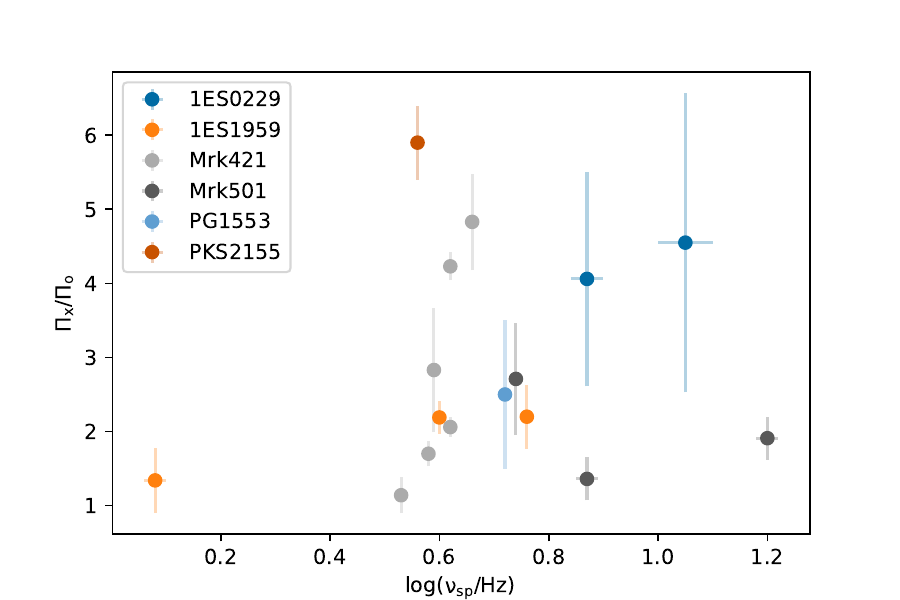}
\end{minipage}
\caption{Plot of the logarithm of the intrinsic 2-10 keV luminosity vs the X-ray polarization degree (left) and the X-ray to optical polarization ratio (right) for the time-averaged observations. Error bars are 68\% confidence intervals.  \label{fig:loglumin}}
\end{figure*}

\begin{figure*}
\centering
\includegraphics[scale=0.50]{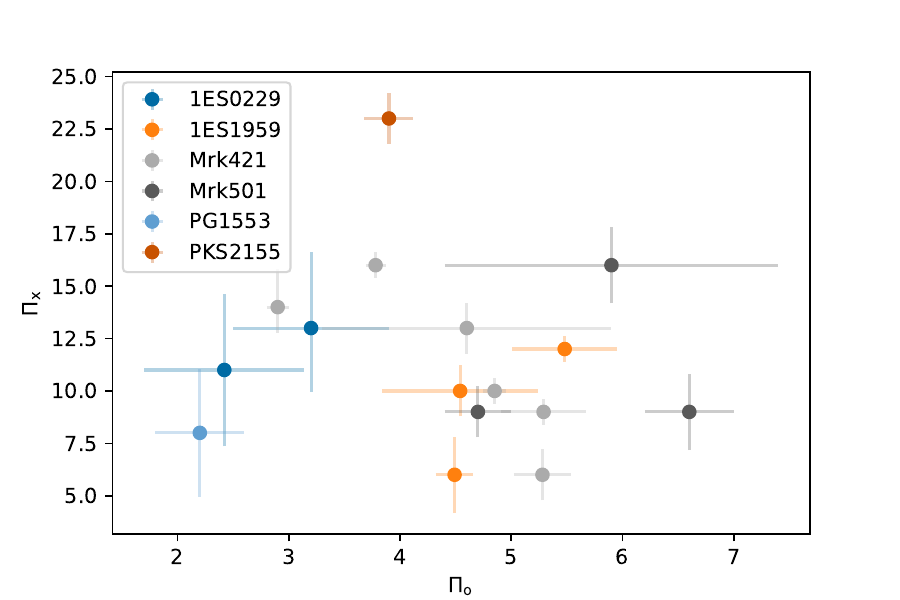}
    \caption{Plot of the the X-ray polarization degree $\Pi_{x}$ vs the optical polarization degree $\Pi_{o}$ for the time averaged observations. Error bars are 68\% confidence intervals  \label{fig:x&o}}
\end{figure*}

We plotted the \texttt{zlogpar} parameters $\alpha$ and $\beta$ , the logarithm of the synchrotron peak frequency, the hardness ratio, and the logarithm of the 2-10 keV luminosity, versus the X-ray polarization degree $\Pi_x$, for the time-averaged sample. We also plot these parameters versus the X-ray to optical polarization degree ratio, $\Pi_x/\Pi_o$. The plots for $\alpha$ are shown for the time averaged sample in Figure \ref{fig:alpha}, whereas the plots for $\beta$ are shown in Figure \ref{fig:beta}. The plots for $\mathrm{log(\nu_{sp}/Hz)}$, the hardness ratio, and $\mathrm{log(L_{2-10})}$ are shown in Figures \ref{fig:sp}, \ref{fig:HR}, and \ref{fig:loglumin}. From Figure \ref{fig:beta} we can see a potential correlation between $\beta$ and $\Pi_x$. Another potential correlation appears when $\beta$ is plotted against $\Pi_x/\Pi_o$. 

\begin{figure*}
\centering
\begin{minipage}{.5\textwidth}
\includegraphics[scale=0.5]{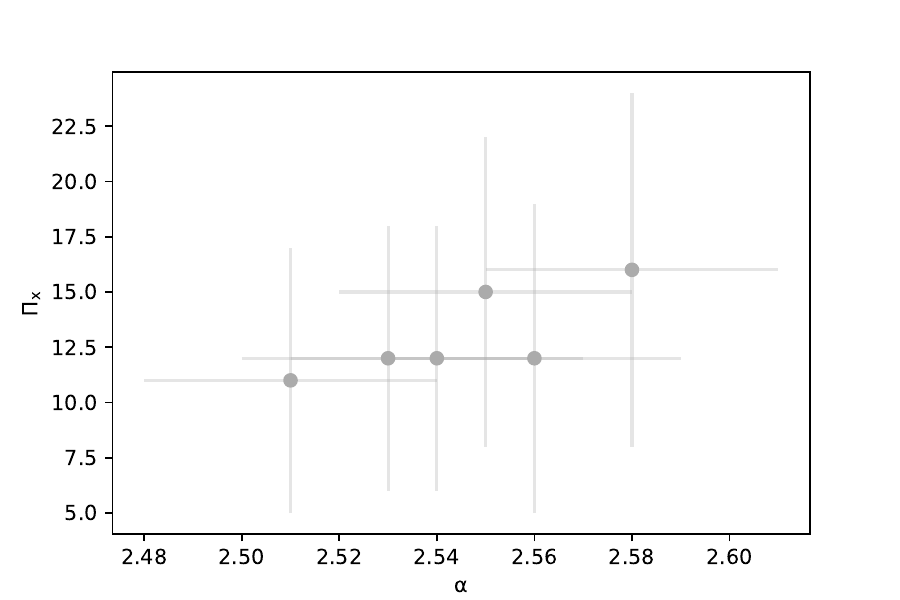}
\end{minipage}%
\begin{minipage}{.5\textwidth}
\includegraphics[scale=0.5]{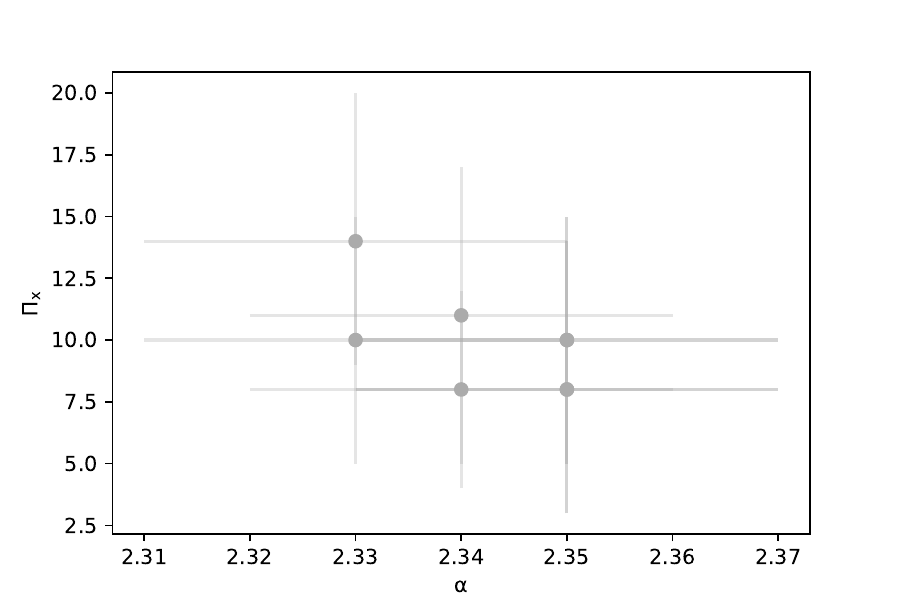}
\end{minipage}
\caption{Plot of the \texttt{zlogpar} parameter $\alpha$ vs the X-ray polarization degree for the time-resolved Mrk 421 spectra from obsID 01003801 (left) and obsID 01003901 (right). Error bars are 90\% confidence intervals. \label{fig:alpha_Mrk421}}
\end{figure*}

\begin{figure*}
\centering
\begin{minipage}{.5\textwidth}
\includegraphics[scale=0.5]{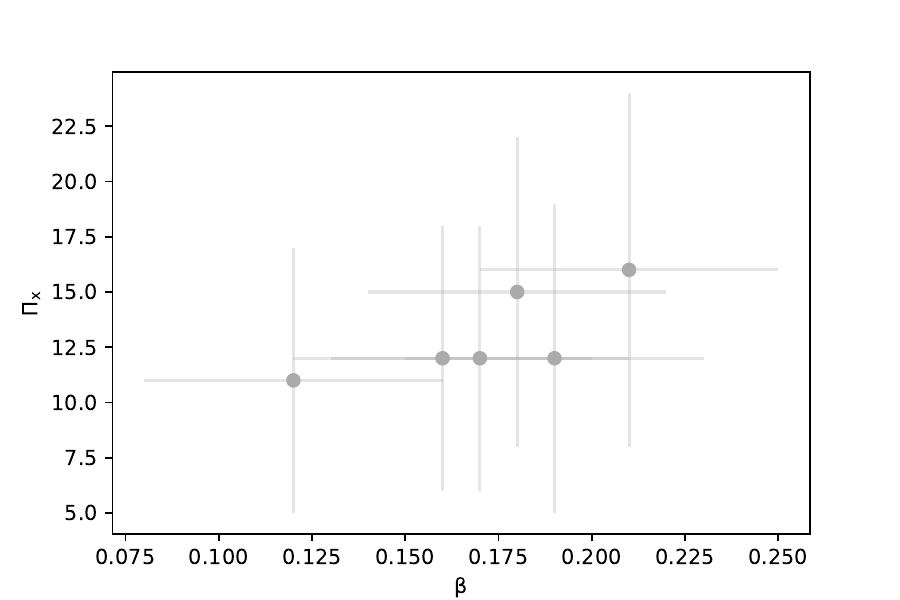}
\end{minipage}%
\begin{minipage}{.5\textwidth}
\includegraphics[scale=0.5]{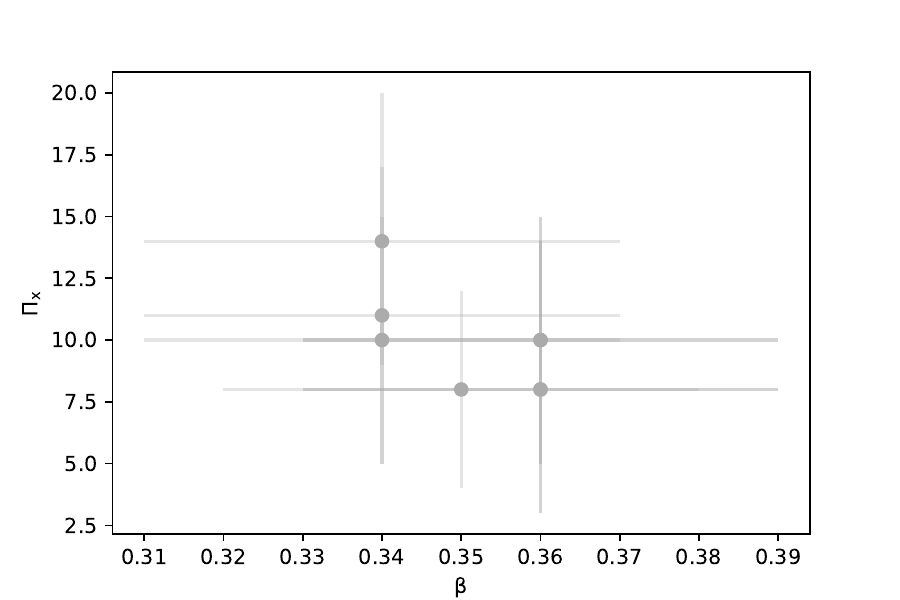}
\end{minipage}
\caption{Plot of the \texttt{zlogpar} parameter $\beta$ vs the X-ray polarization degree for the time-resolved Mrk 421 obsIDs 01003801 (left) and  01003901 (right). Error bars are 90\% confidence intervals.  \label{fig:beta_Mrk421}}
\end{figure*}

\begin{figure*}
\centering
\begin{minipage}{.5\textwidth}
\includegraphics[scale=0.5]{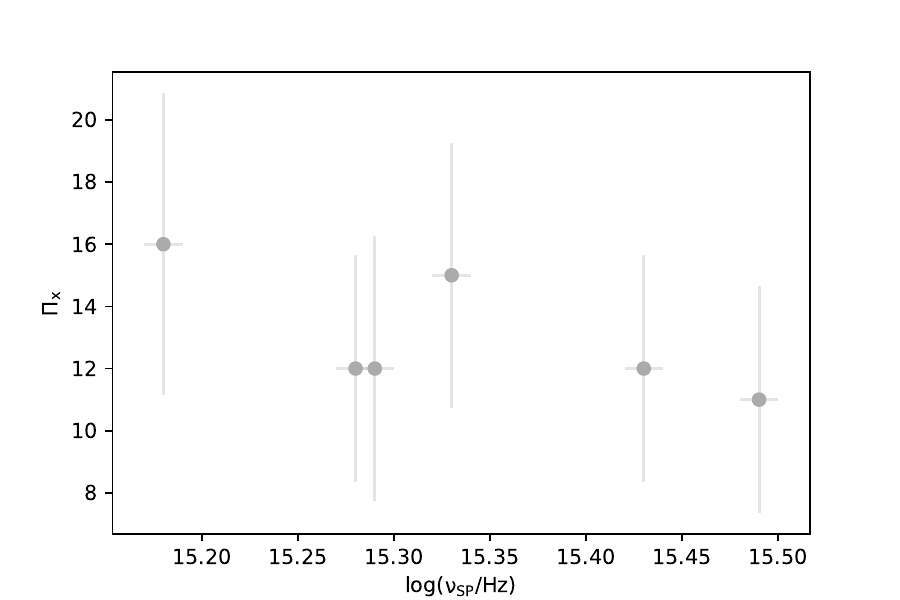}
\end{minipage}%
\begin{minipage}{.5\textwidth}
\includegraphics[scale=0.5]{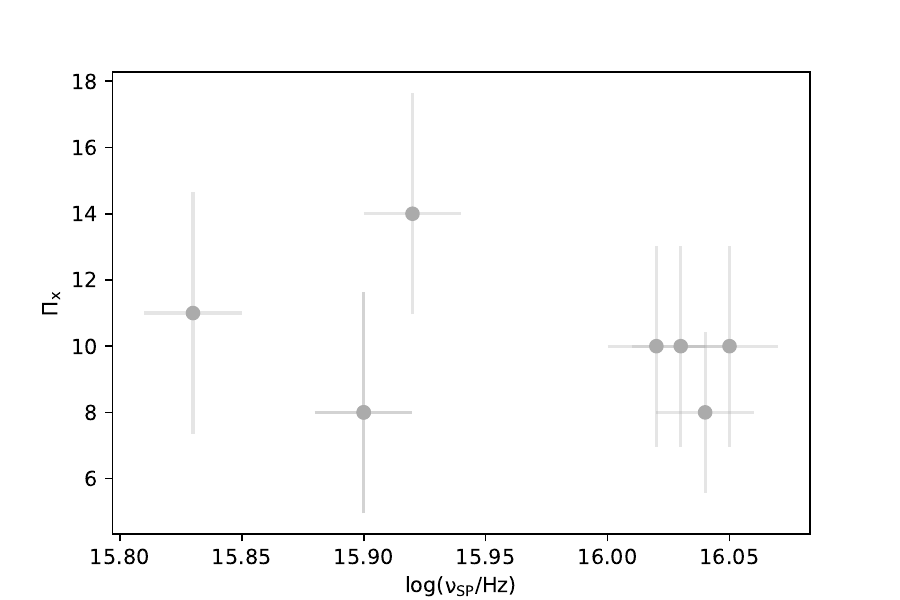}
\end{minipage}
\caption{Plot of the logarithm of the synchrotron peak frequency vs the X-ray polarization degree for the time-resolved Mrk 421 obsIDs 01003801 (left) and  01003901 (right). Error bars are 68\% confidence intervals.  \label{fig:sp_Mrk421}}
\end{figure*}

\begin{figure*}
\centering
\begin{minipage}{.5\textwidth}
\includegraphics[scale=0.5]{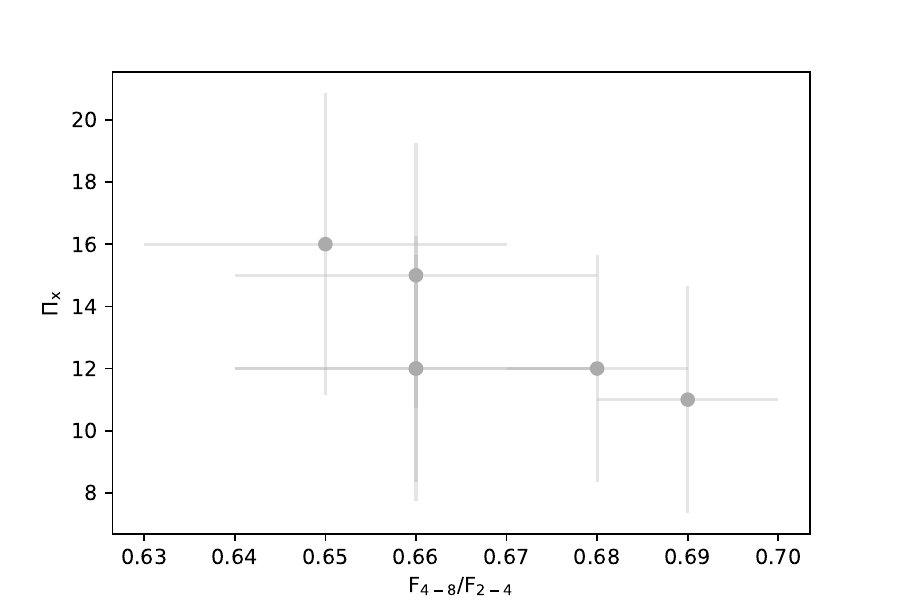}
\end{minipage}%
\begin{minipage}{.5\textwidth}
\includegraphics[scale=0.5]{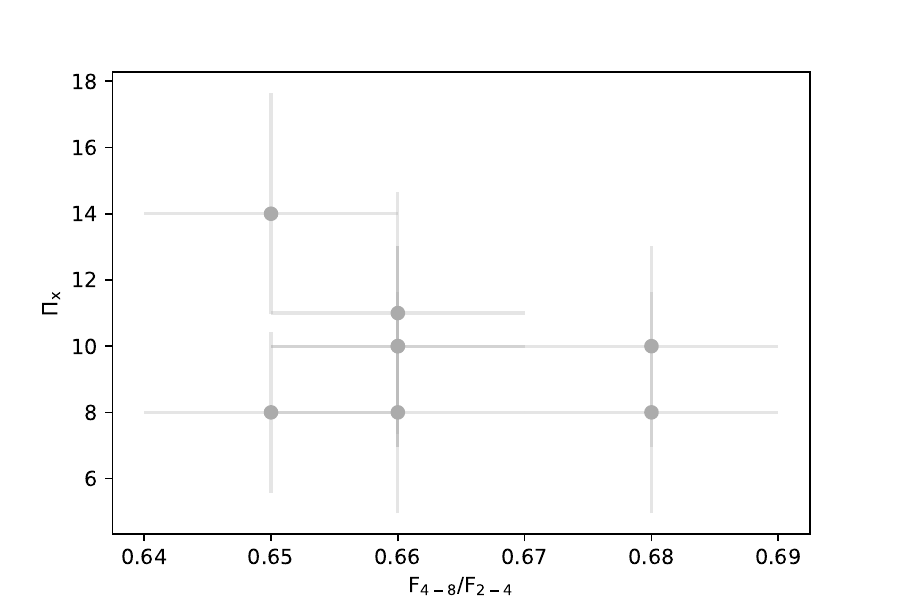}
\end{minipage}
\caption{Plot of the hardness ratio vs the X-ray polarization degree for the time-resolved Mrk 421 obsIDs 01003801 (left) and 01003901 (right). Error bars are 68\% confidence intervals.
\label{fig:HR_Mrk421}}
\end{figure*}

\begin{figure*}
\centering
\begin{minipage}{.5\textwidth}
\includegraphics[scale=0.5]{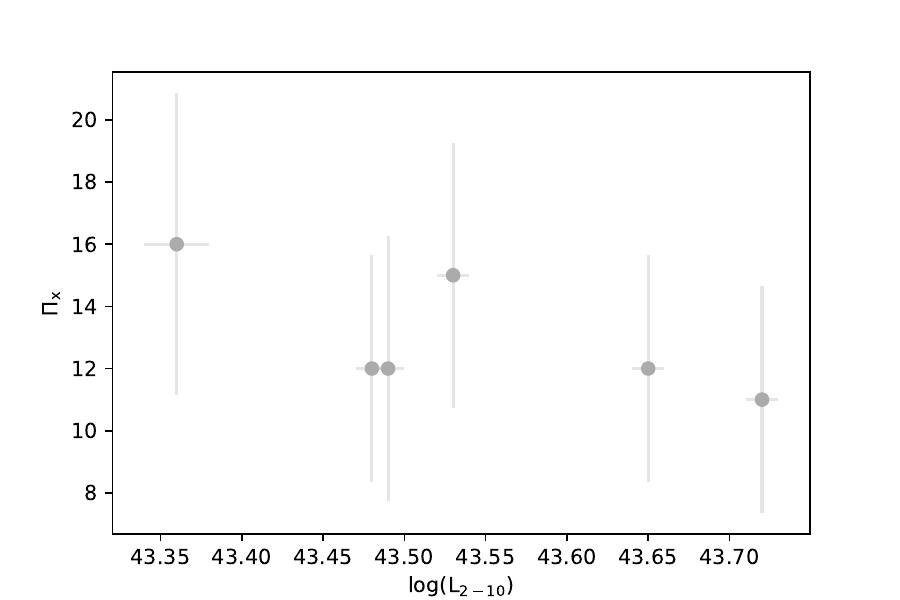}
\end{minipage}%
\begin{minipage}{.5\textwidth}
\includegraphics[scale=0.5]{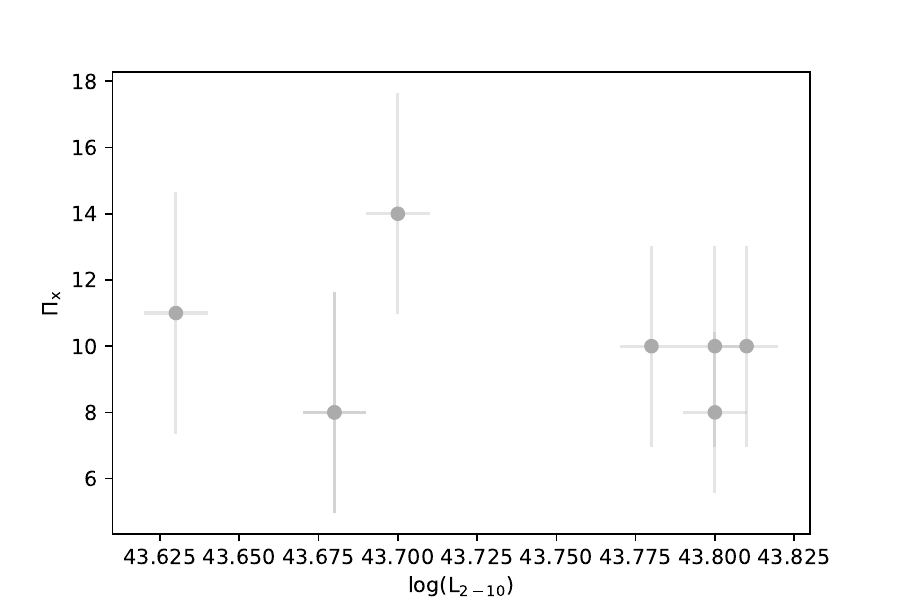}
\end{minipage}
\caption{Plot of the logarithm of the intrinsic 2-10 keV luminosity vs the X-ray polarization degree for the time-resolved Mrk 421 obsIDs 01003801 (left) and 01003901 (right). Error bars are 68\% confidence intervals.\label{fig:loglumin_Mrk421}}
\end{figure*}

These plots do not include Mrk 421 obsIDs 01003801 and 01003901. We plot each spectral parameter vs $\mathrm{\Pi_x}$ separately for these observations in Figures \ref{fig:alpha_Mrk421}, \ref{fig:beta_Mrk421}, \ref{fig:sp_Mrk421}, \ref{fig:HR_Mrk421}, and \ref{fig:loglumin_Mrk421}. The 01003801 data show potential correlations between $\alpha$, $\beta$, and $\Pi_x$, while a potential correlation between the hardness ratio and $\Pi_x$ is visible for 01003901.

We plot $\Pi_x$ against the optical polarization degree $\Pi_o$ for the time-averaged sample in Figure \ref{fig:x&o}. The two quantities do not seem to be obviously correlated. We omit a similar plot for the time-resolved Mrk 421 spectra as the optical polarization degree was assumed to be constant during each observation.

To test the strength of the correlations between each spectral variable and the polarization, we used the Spearman's rank correlation coefficients. The coefficients and resulting p-values are shown for the entire sample in Table \ref{tab:corr}, and for the time-resolved Mrk 421 spectra specifically in Tables \ref{tab:corr_Mrk421obs2} and \ref{tab:corr_Mrk421obs3}. None of the correlations appear to be statistically significant for the entire sample. This is still true when using only the time-averaged data, though the their correlation coefficients and p-values are not shown here. However, when looking at the Mrk 421 01003801 data alone, statistically significant correlations are found between $\alpha$, $\beta$, the hardness ratio, and $\Pi_x$. The Spearman coefficient itself does not account for the error bars, so these potential correlations should be taken with a grain of caution. To take into account the error bars, we ran 100,000 Monte Carlo simulations for each pair of variables, drawing the values of the variables from a random Gaussian distribution with $\sigma$ equal to the width of the 68\% confidence error bars. If the error bars were uneven, we used the upper error bar as the value of $\sigma$. In the case of variables originally measured with 90\% confidence intervals, we converted the error bars into 68\% confidence intervals by dividing the original error bars by 1.645. We then ran the Spearman test on each run.  Including all the data, the strongest potential correlation was $\alpha$ vs $\Pi_x$, which had only 4.04\% of trials result in a p-value of less than 0.05. We also performed Monte Carlo trials on the Mrk 421 time-resolved observations separately. The strongest correlations, $\alpha$, $\beta$, and the hardness ratio for obsID 01003801, only had about 8\% of the trials result in a p-value of less than 0.05.

We also performed the above correlation tests for both Mrk 421 obsIDs 01003801 and 01003901 combined together. These results are shown in Table \ref{tab:corr_Mrk421all}. Similarly to the results for Mrk 421 01003801 alone, multiple statistically significant correlations are found when the error bars are not taken into account. However, none of the correlations gave a statistically significant result in the majority of the Monte Carlo tries. So no statistically significant correlations are found for the time-resolved observations once error bars are taken into account. This is likely due to the large size of the error bars on the parameters relative to the scatter in the nominal values, as seen for example in Figure \ref{fig:alpha_Mrk421}.

\section{Discussion \& Conclusions}\label{sec:discussion} 

From our work it can be seen that there is no definitive evidence for a correlation between HSPs' X-ray spectroscopic and polarization properties.

In a previous review of the IXPE HSP observations, \citet{kim2024b} report a trend between $\Pi_{x}$ and $\Pi_{x}/\Pi_{o}$, which could be due to a genuine correlation between $\Pi_{x}$ and $\Pi_{o}$, or it could be because $\Pi_{x}/\Pi_{o}$ depends on $\Pi_{x}$. We do not find a statistically significant correlation between $\Pi_x$ and $\Pi_o$, which suggests the trend in \citet{kim2024b} is due to $\Pi_{x}/\Pi_{o}$ depending on $\Pi_{x}$. \citet{kim2024b} also suggest that the ratio of $\Pi_{x}$ to $\Pi_{o}$ increases when the blazar spectrum becomes softer. We do not find a statistically significant correlation between the hardness ratio and $\Pi_{x}/\Pi_{o}$, which implies we do not see evidence that  $\Pi_{x}/\Pi_{o}$ increases as the spectrum becomes softer. It should be noted that the statistical significance of the trend identified in \citet{kim2024b} was not evaluated. We should also note that in this study we have performed a uniform analysis on all HSP IXPE observations published at the time of writing. This is in contrast to \citet{kim2024b} who drew the X-ray parameters from the existing literature.

We do not find a correlation in the X-rays between the synchrotron peak frequency $\nu_{sp}$ and $\Pi_x$, {\bf seemingly in} contrast to \citet{angelakis2016} who found an anti-correlation between the optical polarization degree and the synchrotron peak frequency for gamma-ray loud blazars. Our results are also seemingly in contrast to \citet{lister2011}, who found a similar anti-correlation between the radio PD and the synchrotron peak frequency. 

The proposed explanation for the anti-correlation in the optical is energy stratification, in which the optical band samples the higher energy part of the synchrotron peak in low synchrotron peaked blazars (LSPs), where the emission would be closer to the shock and therefore the polarization would be  higher. In  HSPs, the optical band would sample the lower energy part of the synchrotron peak, where the polarization would be lower. Note, however, for HSPs, X-rays sample the high energy portion of the synchrotron hump, the same region of the hump that optical samples in LSPs. This would make their proposed mechanism for the anti-correlation invalid for X-ray polarizations. Therefore our results are not necessarily inconsistent with \citet{angelakis2016}.

In \citet{lister2011}, it was observed that high synchrotron-peaked blazars have significantly lower radio polariaztions. In both LSPs and HSPs, radio samples the low energy side of the synchrotron hump, so the reason for this anti-correlation might be different than the the reason for the one observed in \citet{angelakis2016}. Looking at Figure 10 in \citet{lister2011}, the anti-correlation between the synchrotron peak and the radio PD is much less apparent when only HSPs are included. The fact we do not find a statistically significant correlation between $\nu_{sp}$ and $\Pi_x$ could therefore also be due to the fact we have only examined HSPs in this paper, whereas \citet{lister2011} included LSPs as well. Our results are therefore not necessarily inconsistent with \citet{lister2011}.

The polarization degree is directly related to the uniformity of the magnetic field at the jet electrons' acceleration site, while the spectral properties are directly related to the properties of the electron energy distribution. The lack of a fully statistically significant correlation between the X-ray polarimetric properties and the X-ray spectral properties implies there is little physical connection between the X-ray emitting electron spectrum and the coherence of the magnetic field at the acceleration site. A similar lack of correlation between the energy output of the jet and the polarization degree has been observed for the most recent IXPE observation of PG 1553+113 \citep{middei2026}. Independence of the electron properties and magnetic field properties is expected to happen in an energy-stratified shock scenario  (Fabrizio Tavecchio, private communication). It is uncertain if it is expected in a magnetic reconnection scenario. In general both scenarios predict very similar spectral properties \citep{baring2017,bottcher2019}, so the lack of correlation does not necessarily favor one acceleration mechanism over the other at this point in time. 

The lack of statistical significance for the tested correlations in our work could also be due to the small sample size investigated. We leave the reanalysis of these potential correlations to future work on a larger sample with additional data. The importance of XMM-Newton and NuSTAR observations simultaneous with future IXPE observations cannot be overstated, as without them the spectral parameters are more difficult to measure, and the larger uncertainties could weaken any potential correlations.

More theoretical work is necessary to understand what direct shock acceleration and magnetic reconnection predict for the relationship between the electron spectrum and the polarization properties of the output X-rays. Models often inject an electron distribution into the jet with a pre-determined power law index \citep[e.g.,][]{zhang2014,marscher2014,tavecchio2020}, which could make it hard for them to test the dependence of the polarization on the electron spectrum. Our blazar photon spectra are fit well by a log-parabolic model, but no theoretical simulations have assumed a log-parabolic distribution of electron energies that would be necessary to produce this spectral shape. Our results motivate future theoretical work that can model more complex electron distributions, eventually from first principles, and could help guide the development of models that can predict the correlation between the X-ray polarization and the electron distribution.

\section*{Acknowledgements}

The Imaging X-ray Polarimetry Explorer (IXPE) is a joint US and Italian mission.  The US contribution is supported by the National Aeronautics and Space Administration (NASA) and led and managed by its Marshall Space Flight Center (MSFC), with industry partner Ball Aerospace (now, BAE Systems).  The Italian contribution is supported by the Italian Space Agency (Agenzia Spaziale Italiana, ASI) through contract ASI-OHBI-2022-13-I.0, agreements ASI-INAF-2022-19-HH.0 and ASI-INFN-2017.13-H0, and its Space Science Data Center (SSDC) with agreements ASI-INAF-2022-14-HH.0 and ASI-INFN 2021-43-HH.0, and by the Istituto Nazionale di Astrofisica (INAF) and the Istituto Nazionale di Fisica Nucleare (INFN) in Italy.  This research used data products provided by the {\it IXPE} Team (MSFC, SSDC, INAF, and INFN) and distributed with additional software tools by the High-Energy Astrophysics Science Archive Research Center (HEASARC), at NASA Goddard Space Flight Center (GSFC). I.L. and S.C. were funded by the European Union ERC-2022-STG - BOOTES - 101076343. Views and opinions expressed are however those of the author(s) only and do not necessarily reflect those of the European Union or the European Research Council Executive Agency. Neither the European Union nor the granting authority can be held responsible for them.

\section*{Data Availability}

The data underlying this article are in the public domain, and available on the website of  \href{https://heasarc.gsfc.nasa.gov/}{NASA's High Energy Astrophysics Science Archive Research Center}.



\bibliographystyle{mnras}
\bibliography{refs} 

\bsp	
\label{lastpage}
\end{document}